\documentclass[aps,prd,preprint,nofootinbib,11pt]{revtex4}
\usepackage{amsfonts}
\usepackage{mathrsfs}
\usepackage{graphicx}
\usepackage{amsmath}
\usepackage{amssymb}
\usepackage{subfigure}
\usepackage{epsfig}
\usepackage{graphicx}
\usepackage{color}
\newcommand{\bqa}{\begin{eqnarray}}
\newcommand{\eqa}{\end{eqnarray}}
\newcommand{\beq}{\begin{equation}}
\newcommand{\eeq}{\end{equation}}
\allowdisplaybreaks[1]
\graphicspath{{fig/}{dia/}} \DeclareGraphicsExtensions{.eps}

\begin{document}
\title{Hidden-bottom and -charm hexaquark states in QCD sum rules}
\author{Bing-Dong Wan$^{1}\footnote{wanbingdong16@mails.ucas.ac.cn}$, Liang Tang$^{2}\footnote{tangl@hebtu.edu.cn}$, Cong-Feng Qiao$^{1,3}\footnote{qiaocf@ucas.ac.cn, corresponding author}$}

\affiliation{$^1$ School of Physics, University of Chinese Academy of Science, Yuquan Road 19A, Beijing 10049 \\
$^2$ College of Physics, Hebei Normal University, Shijiazhuang 050024, China\\
$^3$ CAS Center for Excellence in Particle Physics, Beijing 10049, China}

\author{~\\~\\}

\begin{abstract}
\vspace{0.3cm}
In this paper, we investigate the spectra of the prospective hidden-bottom and -charm hexaquark states with quantum numbers $J^{PC }= 0^{++}$, $0^{-+}$, $1^{++}$ and $1^{--}$ in the framework of QCD sum rules. By constructing appropriate interpreting currents, the QCD sum rules analyses are performed up to dimension 12 of the condensates. Results indicate that there exist two possible baryonium states in $b$-quark sector with masses $11.84 \pm 0.22$ GeV and $11.72 \pm 0.26$ GeV for $0^{++}$  and $1^{--}$, respectively. The corresponding hidden-charm partners are found lying respectively at $5.19 \pm 0.24$ GeV and $4.78 \pm 0.23$ GeV. Note that these baryonium states are all above the dibaryon thresholds, which enables their dominant decay modes could be measured at BESIII, BELLEII, and LHCb detectors.
\end{abstract}
 \maketitle
\newpage

\section{Introduction}
Hadrons with more than the minimal quark content ($q\bar{q}$ or $qqq$) was proposed by Gell-Mann \cite{GellMann:1964nj} and Zweig \cite{Zweig} in 1964, which were named as multiquark exotic states and do not infringe the Quantum Chromodynamics (QCD). Exploring the existence and properties of such exotic states is one of the most intriguing research topics of hadronic physics. In past few decades, research on the heavy-flavor exotic states has made tremendous developments, that is, many charmonium-like/bottomonium-like XYZ states have been observed \cite{Choi:2003ue, Aubert:2005rm, Belle:2011aa, Ablikim:2013mio, Liu:2013dau}. In 2015, two hidden-charm pentaquarks $P_c(4380)$ and $P_c(4450)$ \cite{Aaij:2015tga} were observed by the LHCb Collaboration in the $J/\psi$ invariant mass spectrum via the $\Lambda^0_b \to J/\psi p K^-$ process. Recently, the LHCb Collaboration reported a new narrow state $P_c(4312)$, and the previously observed structure $P_c(4450)$ appears to be split into two narrower structures $P_c(4440)$ and $P_c(4457)$ \cite{Aaij:2019vzc}. The recent experimental and theoretical progresses can be found in comprehensive reviews like \cite{Chen:2016qju, Esposito:2016noz, Guo:2017jvc, Olsen:2017bmm, Brambilla:2019esw}.

Facing the observations of tetraquark and pentaquark states, it is nature to conjecture that there should exist the hidden-charm and -bottom hexaquark states, and it is time to hunt for them. For the hexaquark states, the deuteron is a typical and well-established dibaryon molecular state with $J^P=1^+$ and binding energy $E_B = 2.225 \text{MeV}$ \cite{Weinberg:1962hj}. The baryonium states composed of a baryon and an anti-baryon is another special class of heaxquark configuration. In Refs. \cite{Qiao:2005av, Qiao:2007ce}, the $\Lambda_c$-$\bar{\Lambda}_c$ structure was introduced to explain the production and decays of $Y(4260)$. And the heavy baryonium was explored as well in the heavy baryon chiral perturbation theory \cite{Chen:2011cta, Chen:2013sba}.

The method of QCD sum rules~\cite{Shifman, Reinders:1984sr, Narison:1989aq, P.Col} has been applied successfully to many hadronic phenoemena, such as the hadron spectrum and hadron decays. In QCD sum rules, based on the proper interpreting currents corresponding to a hadron of interest, one can construct the two-point or three-point correlation functions, which are respectively used to evaluate the mass and the decay property of the related hadron. Then by matching its operator product expansion (OPE) to its hadronic saturation, the main function for extracting the mass or decay rate of the hadron is established. Utilizing this approach, several significant researches for the hexaquark states have been done \cite{HuaXing:2015, Chen:2016ymy, Sundu:2019, Chen:2019vdh, Wang:2017}. In Ref. \cite{HuaXing:2015}, six-quark state $d^*(2380)$ was considering as $\Delta$-$\Delta$ structure and the corresponding mass was given. Chen {\it et al.} gave an explicit QCD sum rule investigation for hidden-charm baryonium states with various relevant local interpreting currents, and they found some of these currents can couple to hidden-charm baryonium states with the masses around 5.0 GeV \cite{Chen:2016ymy}. Ref. \cite{Sundu:2019} calculated the mass and coupling constant of the scalar hexaquark $uuddss$. Very recently, the $\Omega\Omega$ bound system spectrum with $J^P=0^+$ and $2^+$ in a molecular picture were investigated in Ref. \cite{Chen:2019vdh}, and the results suggest the existence of two bound $\Omega\Omega$ dibaryon states. Wang \cite{Wang:2017} studied the scalar-diquark-scalar-diquark-scalar-diquark type hexaquark state with the QCD sum rules, where the three diquarks were arranged as $ud$, $uc$, and $dc$, respectively.

Besides the above mentioned hexaquark configurations, the QCD theory allows many other possible hexaquark structures, which can couple to hidden-charm baryonium states. Moreover, it should be noted that exploring the hidden-bottom baryonium states is also significant, which is the main motivation of this work, and they tend to be measurable in the LHCb experiment. We investigate the hidden-bottom molecular states in $\Lambda_Q$-$\bar{\Lambda}_Q$ configuration with quantum numbers $J^{PC }= 0^{++}$, $0^{-+}$, $1^{++}$ and $1^{--}$ in the framework of QCD sum rules. Their decay properties, as well as their hidden-charm partners, are also analyzed.

The rest of the paper is arranged as follows. After the introduction, some primary formulas of the QCD sum rules in our calculation are presented in Sec. \ref{Formalism}. The numerical analysis and results are given in Sec. \ref{Numerical}. In the Sec. \ref{Decay}, possible decay modes of hidden-bottom baryonium states are investigated. The last part is left for conclusions and discussion.

\section{Formalism}\label{Formalism}

The starting point of the QCD sum rules is the two-point correlation function constructed from two hadronic currents with the following form:
\begin{eqnarray}
\Pi(q) &=& i \int d^4 x e^{i q \cdot x} \langle 0 | T \{ j (x)  j^\dagger (0) \} |0 \rangle\,;\\
\Pi_{\mu\nu}(q) &=& i \int d^4 x e^{i q \cdot x} \langle 0 | T \{ j_\mu (x)  j_\nu^\dagger (0) \} |0 \rangle \; ,
\end{eqnarray}
where, $j(x)$ and $j_\mu(x)$ are the relevant hadronic currents with $J = 0$ and 1, respectively.

We use the notion $\eta_{\Lambda_Q}$ to represent the Dirac baryon fields of $\Lambda_Q$ without free Lorentz indices. It was shown in Ref.\cite{Bagan:1992} that, $\eta_{\Lambda_Q}$ may take the following quark structure:
\begin{eqnarray}\label{current_lambda_c}
\eta_{\Lambda_Q}(x)&=&i \epsilon_{a b c}[ q_a^T(x) C \gamma_5 q_b^\prime(x) ]Q_c(x) \; ,
\end{eqnarray}
where $Q=b, c$. Therefore, the interpolating currents for $\Lambda_Q$-$\bar{\Lambda}_Q$ baryounium states with quantum numbers $ 0^{++}$, $0^{-+}$,  $1^{++}$, and $1^{--}$ can be respectively constructed as
\begin{eqnarray}\label{current_lambda}
j^{0^{++}}(x)&=&\epsilon_{a b c} \epsilon_{d e f} [\bar{Q}_d (x) Q_c (x)] [q_a^T (x) C \gamma_5 q_b^\prime (x)] [\bar{q}_e^\prime (x) \gamma_5 C \bar{q}_f^T (x)]\;,\label{J0++}\\
j^{0^{-+}}(x)&=& \epsilon_{a b c} \epsilon_{d e f} [\bar{Q}_d (x) \gamma_5 Q_c (x)] [q_a^T (x) C \gamma_5 q_b^\prime (x)] [\bar{q}_e^\prime (x) \gamma_5 C \bar{q}_f^T (x)]\;,\label{J0-+}\\
j^{1^{++}}_\mu(x)&=& \epsilon_{a b c} \epsilon_{d e f} [\bar{Q}_d (x) \gamma_\mu \gamma_5 Q_c (x)] [q_a^T (x) C \gamma_5 q_b^\prime (x)] [\bar{q}_e^\prime (x) \gamma_5 C \bar{q}_f^T (x)]\;, \label{J1++}\\
j^{1^{--}}_\mu(x)&=&\epsilon_{a b c} \epsilon_{d e f} [\bar{Q}_d (x) \gamma_\mu Q_c (x)] [q_a^T (x) C \gamma_5 q_b^\prime (x)] [\bar{q}_e^\prime (x) \gamma_5 C \bar{q}_f^T (x)]\;. \label{J1--}
\end{eqnarray}
Here, $a \cdots f$ denote color indices, $C$ is the charge conjugation matrix, $Q$ represents the heavy quarks,  $q$ stands for the up quark $u$, and $q^\prime$ for the down quark $d$.

The correlation function derived from Eqs. (\ref{J1++}) and (\ref{J1--}) can be expressed as the following Lorentz covariance form
\begin{eqnarray}
\Pi_{\mu\nu}=-(g_{\mu\nu}-\frac{q_\mu q_\nu}{q^2}) \Pi_1(q^2)+\frac{q_\mu q_\nu}{q^2} \Pi_0(q^2)\;,
\end{eqnarray}
where the subscripts $1$ and $0$ denote the quantum numbers of the spin $1$ and $0$ mesons, respectively. However, since the leading term $\Pi_1(q^2)$ is symmetrical and only contains the spin 1 component, we shall focus on calculating the $\Pi_1(q^2)$ and employ it to perform the  QCD sum rule analyses.

On the phenomenological side, adopting the usual pole plus continuum parametrization of the hadronic the spectral density, we express the correlation function $\Pi(q^2)$ as
\begin{eqnarray}
\Pi^{PHEN}(q^2) & = & \frac{(\lambda^X)^2}{(M^X)^2 - q^2} + \frac{1}{\pi} \int_{s_0}^\infty d s \frac{\rho(s)}{s - q^2} \; , \label{hadron}
\end{eqnarray}
where the superscript $X$ denotes the lowest lying  $\Lambda_Q$-$\bar{\Lambda}_Q$ hexaquark state, $M^{X}$ is its mass, and $\rho(s)$ is the spectral density that contains the contributions from higher excited states and the continuum states above the threshold $s_0$. The decay constant $\lambda^X$ is defined through $\langle 0 | j | X \rangle = \lambda^X $ and $\langle 0 | j_\mu | X \rangle = \lambda^X \epsilon_\mu$.

On the OPE side, the correlation function $\Pi(q^2)$ can be written as a dispersion relation form:
 \begin{eqnarray}
\Pi^{OPE} (q^2) &=& \int_{(2 m_Q + 2m_q + 2m_{q^\prime})^2}^{\infty} d s
\frac{\rho^{OPE}(s)}{s - q^2}\; ,
\label{OPE-hadron}
\end{eqnarray}
where $\rho^{OPE}(s) = \text{Im} [\Pi^{OPE}(s)] / \pi$ is the spectral density of the OPE side, and contains the contributins of the condensates up to dimension 12, thus
\begin{eqnarray}
\rho^{OPE}(s) & = & \rho^{pert}(s) + \rho^{\langle \bar{q} q
\rangle}(s) +\rho^{\langle G^2 \rangle}(s) + \rho^{\langle \bar{q} G q \rangle}(s)
+ \rho^{\langle \bar{q} q \rangle^2}(s) + \rho^{\langle G^3 \rangle}(s) \nonumber\\
&+& \rho^{\langle \bar{q} q \rangle\langle \bar{q} G q \rangle}(s) +  \rho^{\langle \bar{q} q
\rangle^2\langle G^2 \rangle}(s) + \rho^{\langle \bar{q} G q \rangle^2}(s) + \rho^{\langle \bar{q} q
\rangle^4}(s)  \; . \label{rho-OPE}
\end{eqnarray}

\begin{figure}
\begin{center}
\includegraphics[width=10.8cm]{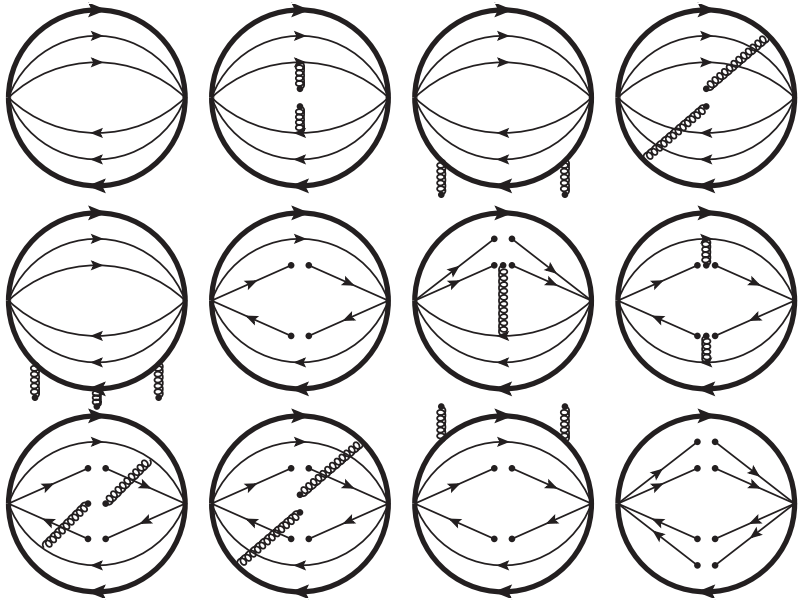}
\caption{The typical Feynman diagrams related to the $\Pi(q^2)$ function, where the thick solid line represents the heavy quark, the thin solid line stands for the light quark, and the spiral line denotes the gluon. There is no heavy quark consendsate due to the large heavy quark mass.} \label{feyn}
\end{center}
\end{figure}

In order to calculate the spectral density of the OPE side, Eq. (\ref{rho-OPE}), the full propagators $S^q_{i j}(x)$ and $S^Q_{i j}(p)$ of a light quark ($q=u$, $d$ or $s$) and a heavy quark ($Q=c$ or $b$) are used:
\begin{eqnarray}
S^q_{j k}(x) \! \! & = & \! \! \frac{i \delta_{j k} x\!\!\!\slash}{2 \pi^2
x^4} - \frac{\delta_{jk} m_q}{4 \pi^2 x^2} - \frac{i t^a_{j k} G^a_{\alpha\beta}}{32 \pi^2 x^2}(\sigma^{\alpha \beta} x\!\!\!\slash
+ x\!\!\!\slash \sigma^{\alpha \beta}) - \frac{\delta_{jk}}{12} \langle\bar{q} q \rangle + \frac{i\delta_{j k}
x\!\!\!\slash}{48} m_q \langle \bar{q}q \rangle - \frac{\delta_{j k} x^2}{192} \langle g_s \bar{q} \sigma \cdot G q \rangle \nonumber \\ &+& \frac{i \delta_{jk} x^2 x\!\!\!\slash}{1152} m_q \langle g_s \bar{q} \sigma \cdot G q \rangle - \frac{t^a_{j k} \sigma_{\alpha \beta}}{192}
\langle g_s \bar{q} \sigma \cdot G q \rangle
+ \frac{i t^a_{jk}}{768} (\sigma_{\alpha \beta} x \!\!\!\slash + x\!\!\!\slash \sigma_{\alpha \beta}) m_q \langle
g_s \bar{q} \sigma \cdot G q \rangle \;,
\end{eqnarray}
\begin{eqnarray}
S^Q_{j k}(p) \! \! & = & \! \! \frac{i \delta_{j k}(p\!\!\!\slash + m_Q)}{p^2 - m_Q^2} - \frac{i}{4} \frac{t^a_{j k} G^a_{\alpha\beta} }{(p^2 - m_Q^2)^2} [\sigma^{\alpha \beta}
(p\!\!\!\slash + m_Q)
+ (p\!\!\!\slash + m_Q) \sigma^{\alpha \beta}] \nonumber \\ &+& \frac{i\delta_{jk}m_Q  \langle g_s^2 G^2\rangle}{12(p^2 - m_Q^2)^3}\bigg[ 1 + \frac{m_Q (p\!\!\!\slash + m_Q)}{p^2 - m_Q^2} \bigg] \nonumber \\ &+& \frac{i \delta_{j k}}{48} \bigg\{ \frac{(p\!\!\!\slash +
m_Q) [p\!\!\!\slash (p^2 - 3 m_Q^2) + 2 m_Q (2 p^2 - m_Q^2)](p\!\!\!\slash + m_Q) }{(p^2 - m_Q^2)^6}
\bigg\} \langle g_s^3 G^3 \rangle \; ,
\end{eqnarray}
where, the vacuum condensates are clearly displayed. For more explanation on above propagators, readers may refer to Refs.~\cite{Wang:2013vex, Albuquerque:2013ija}. The Feynman diagrams corresponding to each term of Eq.(\ref{OPE-hadron}) are schematically shown in Fig.(\ref{feyn}).

Applying Borel transform on both Eq. (\ref{hadron}) and Eq. (\ref{OPE-hadron}), using quark-hadron duality, and matching the OPE side with the phenomenological side of the correlation function $\Pi(q^2)$, we can finally reach the main function to extract the mass of the hexaquark state, which reads
\begin{eqnarray}
M^X(s_0, M_B^2) = \sqrt{- \frac{L_1(s_0, M_B^2)}{L_0(s_0, M_B^2)}} \; . \label{mass-Eq}
\end{eqnarray}
Here $L_1$ and $L_0$ are respectively defined as
\begin{eqnarray}
L_0(s_0, M_B^2) =  \int_{(2m_c + 2 m_q + 2 m_{q^\prime})^2}^{s_0} d s \; \rho^{OPE}(s) e^{-
s / M_B^2}  \;  \label{L0}
\end{eqnarray}
and
\begin{eqnarray}
L_1(s_0, M_B^2) =
\frac{\partial}{\partial{\frac{1}{M_B^2}}}{L_0(s_0, M_B^2)} \; .
\end{eqnarray}

\section{Numerical analysis}\label{Numerical}
In the numerical calculation of QCD sum rules, the values of input parameters we take are \cite{Matheus:2006xi, Cui:2011fj, Narison:2002pw}
\begin{eqnarray}
\begin{aligned}
& m_c (m_c) = (1.23 \pm 0.05) \; \text{GeV} \; , & & m_b (m_b) =
(4.24 \pm 0.06) \; \text{GeV}, \\
& \langle \bar{q} q \rangle = - (0.23 \pm 0.03)^3 \; \text{GeV}^3 \; ,
& & \langle g_s^2 G^2 \rangle = 0.88 \; \text{GeV}^4 \; , \\
& \langle \bar{q} g_s \sigma \cdot G q \rangle = m_0^2 \langle
\bar{q} q \rangle \; , & & \langle g_s^3 G^3 \rangle = 0.045 \;
\text{GeV}^6 \; ,\\
& m_0^2 = 0.8 \; \text{GeV}^2\; , & &
\end{aligned}
\end{eqnarray}
in which the $\overline{\text{MS}}$ running heavy quark masses are adopted. For light quarks $u$ and $d$, we use the chiral limit in our analysis in which their masses are $m_u=m_d=0$.

The QCD sum rules in Eq.(\ref{mass-Eq}) is a function of the Borel parameter $M_B^2$ and the continuum threshold $s_0$. To obtain a reliable mass sum rules result, one should choose suitable working ranges for these two parameters. As widely used, we employ two criteria to fix the suitable working regions of $M_B^2$ and $s_0$ \cite{Shifman,Reinders:1984sr, P.Col}. The first one asks for the convergence of the OPE, which is to compare the relative contribution of each term to the total contribution on the OPE side. The other criterion of QCD sum rules is the pole contribution (PC). As discussed in Ref. \cite{HuaXing:2015,Sundu:2019,Wang:2017}, since the large power of $s$ in the spectral density suppress the PC value, we choose the pole contribution larger than $15\%$ for hexaquark states.

In order to determine a proper value for $s_0$, we carry out a similar analysis in Refs. \cite{Qiao-Tang:2014a}. Since the continuum threshold $s_0$ relates to the mass of the ground state by $\sqrt{s_0} \sim M^X + 0.5 \, \text{GeV}$ \cite{P.Col,Finazzo:2011he}, where $M^X$ denotes the mass of the ground state, various $\sqrt{s_0}$ satisfying this constraint are taken into account. Among these values, one needs to select out the one which yields an optimal window for Borel parameter $M_B^2$. That is, within the optimal window, the hexaquark mass $M^X$ is somehow independent of the Borel parameter $M_B^2$ as much as possible. In practice, in QCD sum rules calculation, we may vary $\sqrt{s_0}$ by $0.2$ GeV, which gives the lower and upper bounds hence the uncertainties of $\sqrt{s_0}$.

\begin{figure}
\begin{center}
\includegraphics[width=6.8cm]{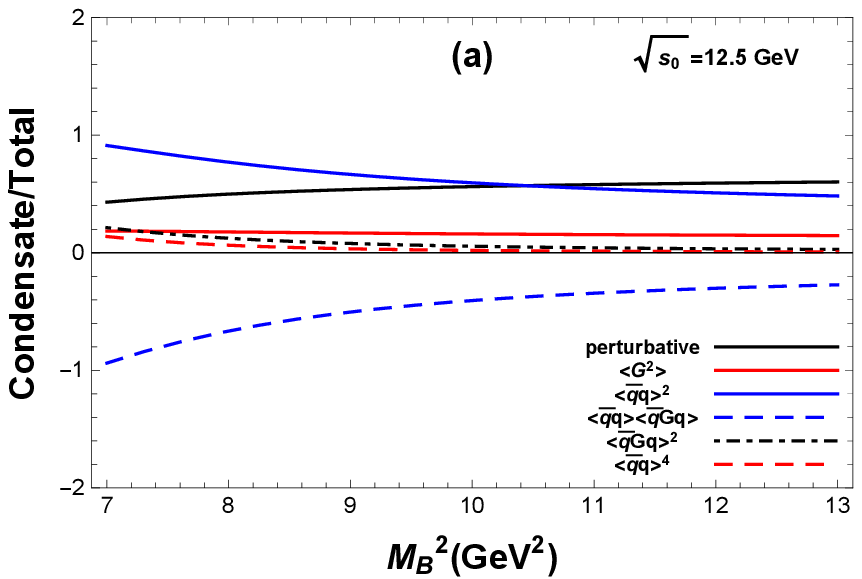}
\includegraphics[width=6.8cm]{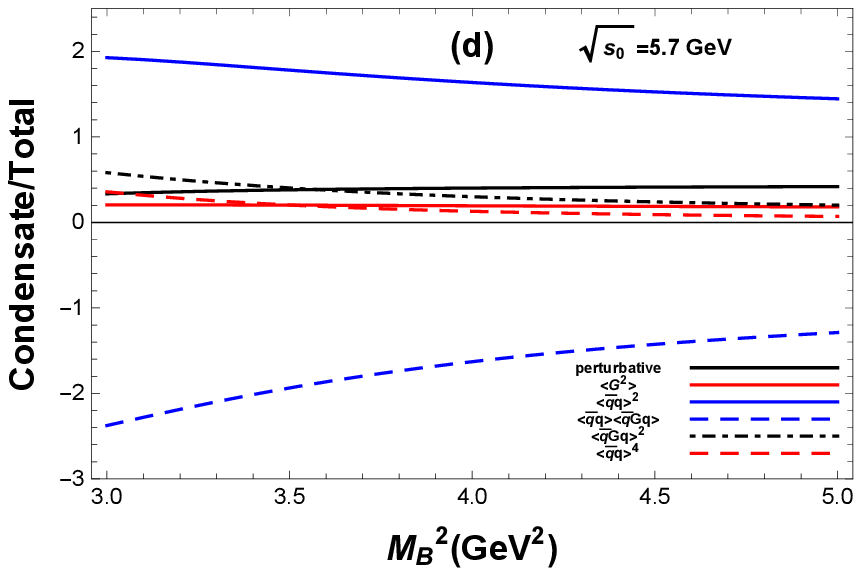}
\includegraphics[width=6.8cm]{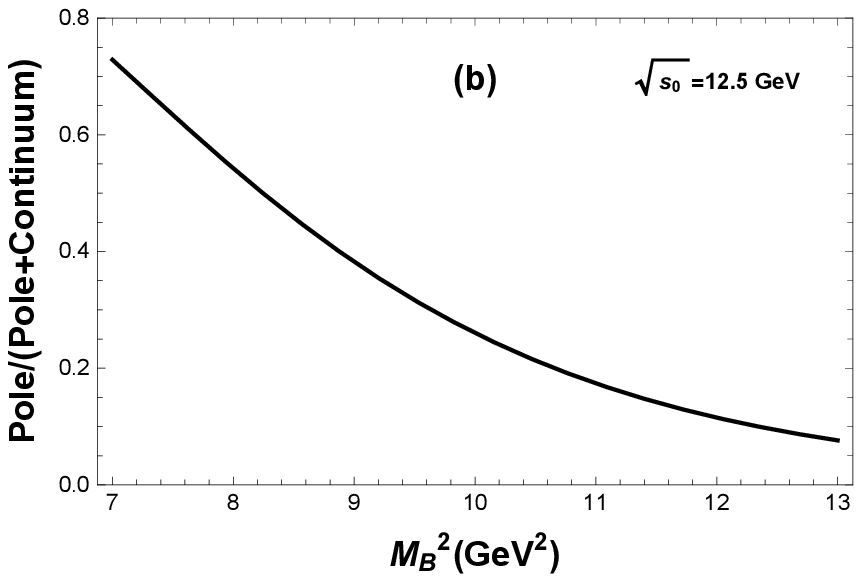}
\includegraphics[width=6.8cm]{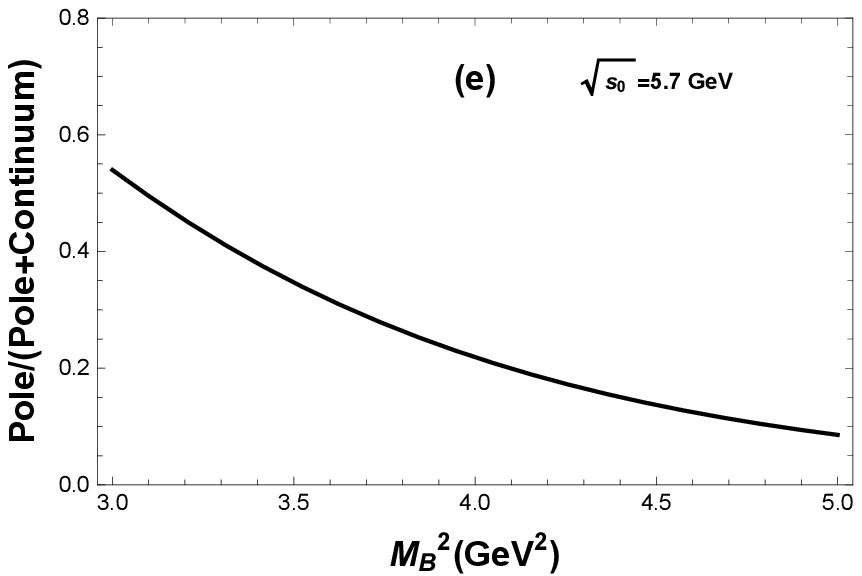}
\includegraphics[width=6.8cm]{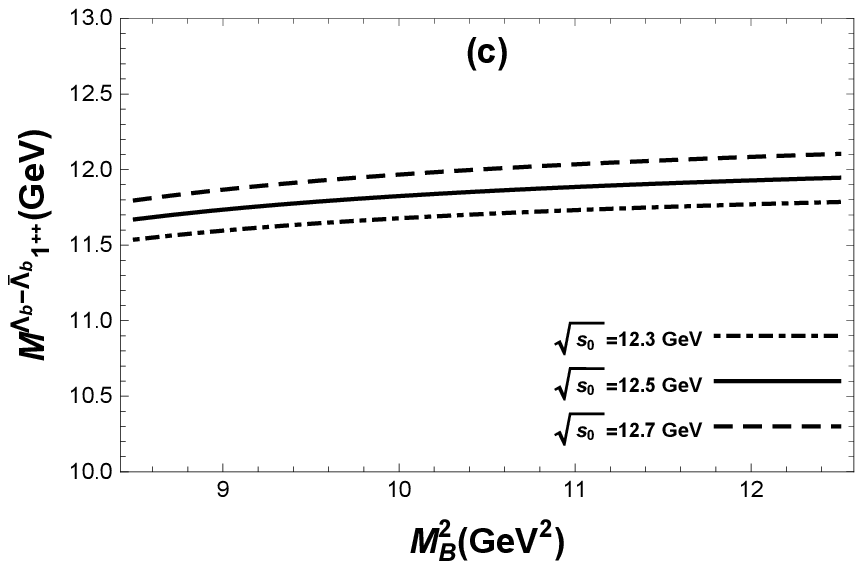}
\includegraphics[width=6.8cm]{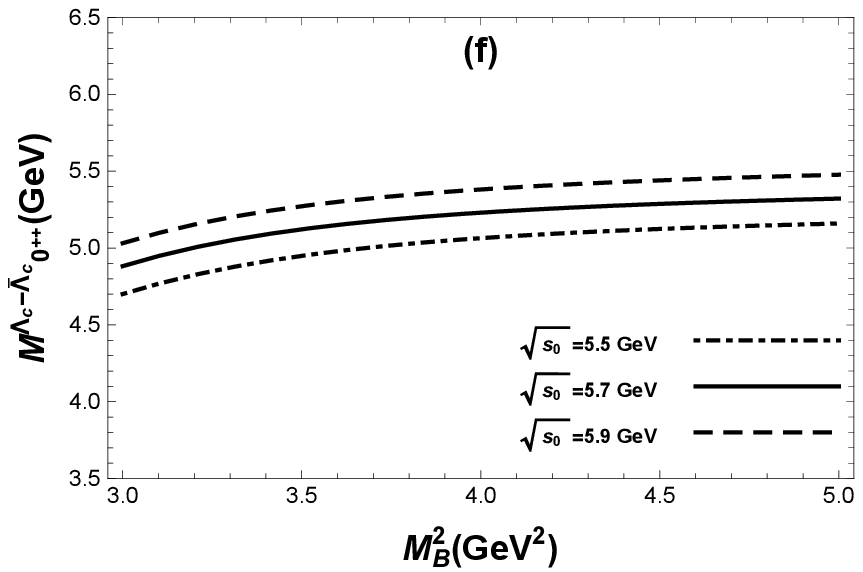}
\caption{Figures for $0^{++}$  hidden-bottom and -charm hexaquark molecular states. (\textbf{a}) The OPE convergence of  hidden-bottom state as function of the Borel parameter $M_B^2$ in the region $7.0 \leq M_B^2 \leq 13.0 \; \text{GeV}^{2}$ with $\sqrt{s_0} = 12.5 \; \text{GeV}$. The $\langle G^3 \rangle$ and $\langle \bar{q} q\rangle^2\langle G^2 \rangle$ terms contributions are not displayed, since their magnitudes are tiny, and $\langle \bar{q} q\rangle$ and $\langle \bar{q} G q\rangle$ are $0$. (\textbf{b}) The pole contribution of hidden-bottom state as the function of the Borel parameter $M_B^2$ with $\sqrt{s_0} = 12.5 \; \text{GeV}$. (\textbf{c}) The mass  of  hidden-bottom hexaquark molecular state $M^{\Lambda_b-\bar{\Lambda}_b}_{0^{++}}$ as the function of the Borel parameter $M_B^2$. (\textbf{d}) The OPE convergence of hidden-charm state as the function of the Borel parameter $M_B^2$ in the region $3.0 \leq M_B^2 \leq 5.0 \; \text{GeV}^{2}$ with $\sqrt{s_0} = 5.7 \; \text{GeV}$.  (\textbf{e}) The pole contribution of hidden-charm state as the function of the Borel parameter $M_B^2$ with $\sqrt{s_0} = 5.7 \; \text{GeV}$. (\textbf{f}) The mass of hidden-charm hexaquark molecular state $M^{\Lambda_c-\bar{\Lambda}_c}_{0^{++}}$ as the function of the Borel parameter $M_B^2$.} \label{fig1}
\end{center}
\end{figure}

After scanning the values of $\sqrt{s_0}$, we can obtain the optimal $\sqrt{s_0}$ together with the suitable window of $M_B^2$. Quantitatively, the OPE convergence of $0^{++}$ hidden-bottom hexaquark state is shown in Fig.(\ref{fig1}-a). According to the first criterion, we find a strong OPE convergence for  $M_B^2\gtrsim9.1\text{GeV}^{2}$ with $\sqrt{s_0}=12.5\, \text{GeV}$, and then we fix the lower working limit for $M_B^2$.  The curve of the PC is illustrated in Fig.(\ref{fig1}-b), which indicates for $PC>15\%$ the upper constraint on $M_B^2$ is $M_B^2 \lesssim 11.4 \; \text{GeV}^{2}$ with $\sqrt{s_0}=12.5\, \text{GeV}$. The masses $M^{\Lambda_b-\bar{\Lambda}_b}_{0^{++}}$ as functions of the Borel parameter $M_B^2$ are drawn in Fig.(\ref{fig1}-c), where the center curve corresponds to the optimal threshold parameter $\sqrt{s_0}$ and the upper (lower) curve is drawn to display the uncertainty on $\sqrt{s_0}$ with $\sqrt{s_0} = 12.7 \text{GeV}$ ($\sqrt{s_0} = 12.3 \text{GeV}$), respectively.

In the charm quark sector, the $0^{++}$  hidden-charm hexaquark state may be analyzed similarly, with heavy quark mass $m_Q = m_c$. The OPE convergence is drawn in Fig.(\ref{fig1}-d), where we find $M_B^2\gtrsim3.5 \text{GeV}^{2}$ with $\sqrt{s_0}=5.7\, \text{GeV}$ is satisfied under the first criterion. On the other hand, the PC is drawn in Fig.(\ref{fig1}-e), which determines the upper bound of $M_B^2$, that is, $M_B^2 \lesssim 4.4 \; \text{GeV}^{2}$ with $\sqrt{s_0}=5.7\, \text{GeV}$ (for $PC>15\%$). Finally, the masses $M^{\Lambda_c-\bar{\Lambda}_c}_{0^{++}}$ as function of the Borel parameter $M_B^2$ for $\sqrt{s_0}= 5.9, 5.7, 5.5 \text{GeV}$ are drawn in Fig.(\ref{fig1}-f), where the upper and lower curves correspond to the uncertainties stemming from $\sqrt{s_0}$.

Finally, the mass spectra of the $0^{++}$  $\Lambda_Q$-$\bar{\Lambda}_Q$ hexaquark molecular states are determined to be
\begin{eqnarray}
M_{0^{++}}^{\Lambda_b-\bar{\Lambda}_b}&=&(11.84\pm0.22)\;\text{GeV}\;,\\
M_{0^{++}}^{\Lambda_c-\bar{\Lambda}_c}&=&(5.19\pm0.24)\;\text{GeV}\,.
\end{eqnarray}
where the errors stems from the uncertainties of the quark masses, the condensates and the threshold parameter $\sqrt{s_0}$.

\begin{figure}
\begin{center}
\includegraphics[width=6.8cm]{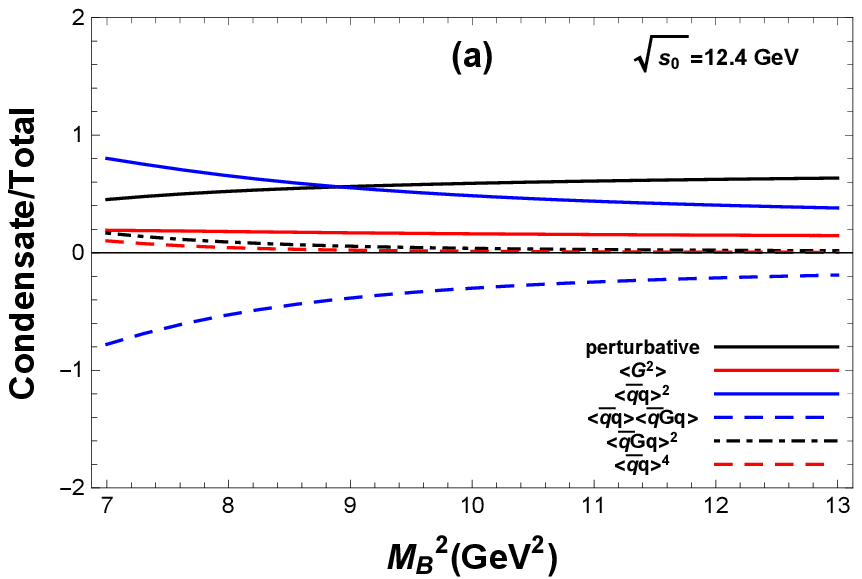}
\includegraphics[width=6.8cm]{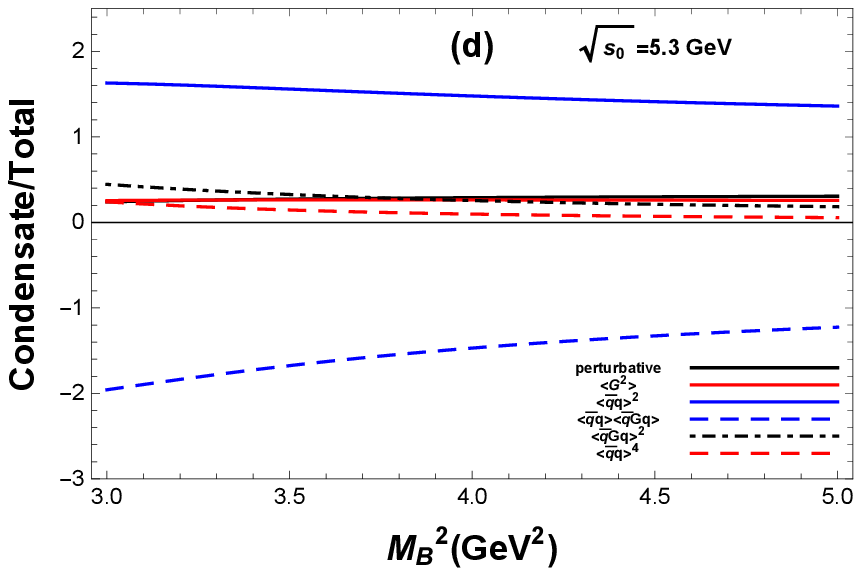}
\includegraphics[width=6.8cm]{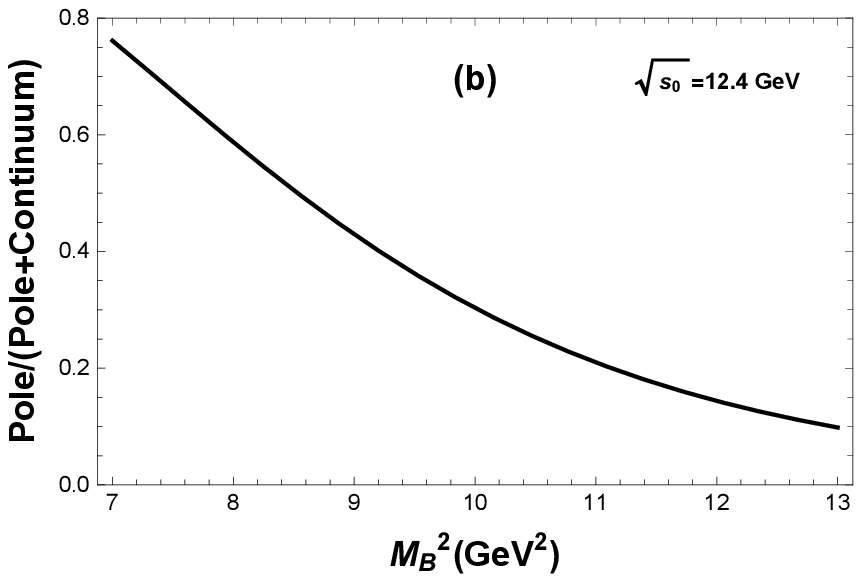}
\includegraphics[width=6.8cm]{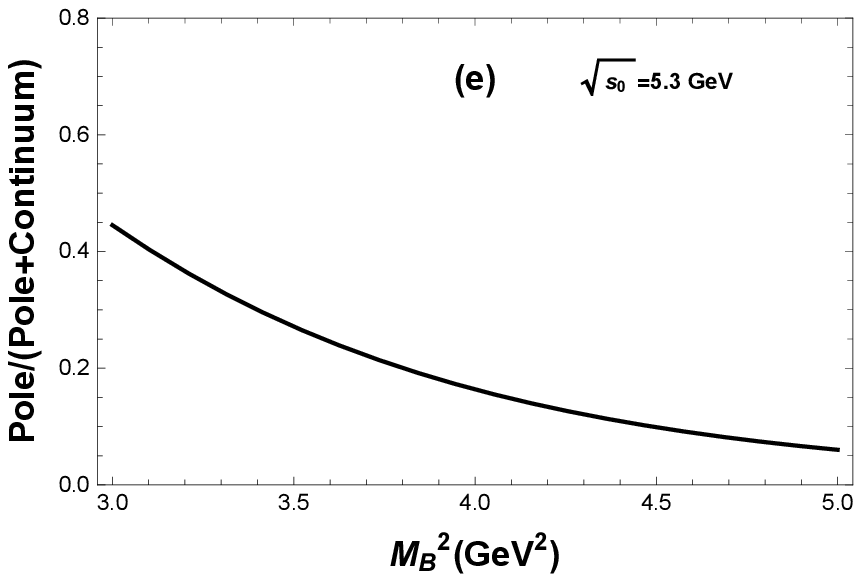}
\includegraphics[width=6.8cm]{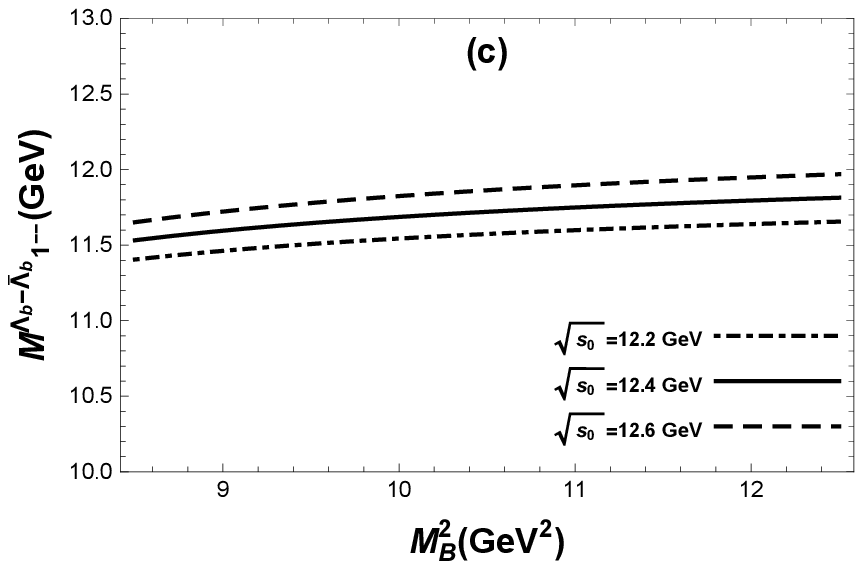}
\includegraphics[width=6.8cm]{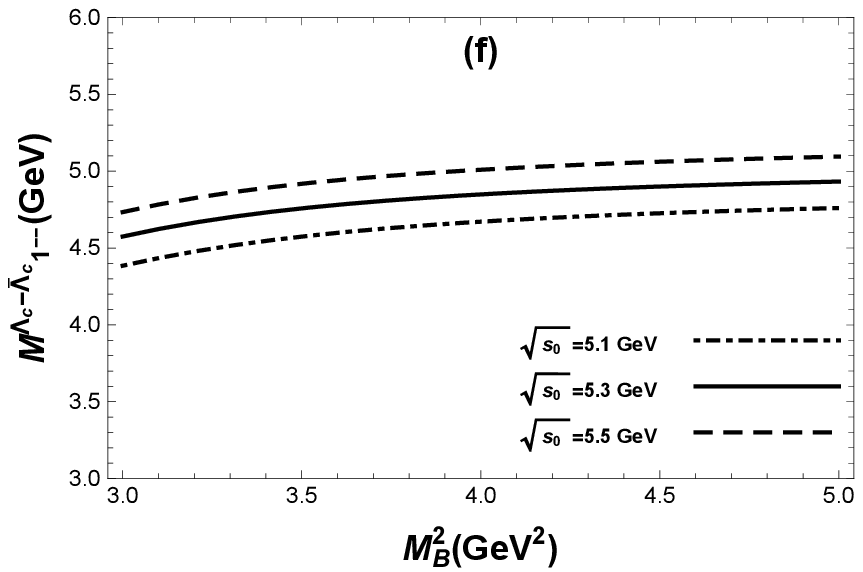}
\caption{The same caption as in Fig.\ref{fig1}, but for $1^{--}$ hidden-bottom and -charm hexaquark molecular states, where continuum thresholds $\sqrt{s_0}$ are taken as $12.20$, $12.40$ and $12.60$ GeV in (\text{c}) for the hidden-bottom case, and as $5.10$, $5.30$, and $5.50$ GeV in (\text{f}) for the hidden-charm case, from down to up.} \label{fig2}
\end{center}
\end{figure}

The OPE convergence of $1^{--}$  hidden-bottom hexaquark state is shown in Fig.(\ref{fig2}-a). Here, we find a strong OPE convergence for  $M_B^2\gtrsim8.9 \text{GeV}^{2}$ with $\sqrt{s_0}=12.4\, \text{GeV}$, which consist in a good criterion for fixing the value of the lower limit of $M_B^2$. The curve of the PC is shown in Fig.(\ref{fig2}-b), which determines the upper bound for $M_B^2$, i.e., $M_B^2 \lesssim 11.9 \; \text{GeV}^{2}$ with $\sqrt{s_0}=12.4.\, \text{GeV}$. The masses $M^{\Lambda_b-\bar{\Lambda}_b}_{1^{--}}$ as functions of the Borel parameter $M_B^2$ for different $\sqrt{s_0}$ are drawn in Fig.(\ref{fig2}-c).

For the $1^{--}$  hidden-charm hexaquark state, the OPE convergence is drawn in Fig.(\ref{fig2}-d). According to the first criterion, we find the lower limit of $M_B^2$, i.e., $M_B^2\gtrsim 3.3 \text{GeV}^{2}$ with $\sqrt{s_0}=5.3\, \text{GeV}$. The PC is drawn in Fig.(\ref{fig2}-e), which fix the upper bound for $M_B^2$, i.e., $M_B^2 \lesssim 4.1 \; \text{GeV}^{2}$ with $\sqrt{s_0}=5.3\, \text{GeV}$. The masses $M^{\Lambda_c-\bar{\Lambda}_c}_{1^{--}}$ as functions of the Borel parameter $M_B^2$ for different $\sqrt{s_0}$ are drawn in Fig.(\ref{fig2}-f).

Eventually, the masses of the $1^{--}$  $\Lambda_Q$-$\bar{\Lambda}_Q$ hexaquark molecular states are determined to be
\begin{eqnarray}
M_{1^{--}}^{\Lambda_b-\bar{\Lambda}_b}&=&(11.72\pm0.26)\;\text{GeV}\;,\\
M_{1^{--}}^{\Lambda_c-\bar{\Lambda}_c}&=&(4.78\pm0.23)\;\text{GeV}\,.
\end{eqnarray}
Here the errors stems from the uncertainties of the quark masses, the condensates and the threshold parameter $\sqrt{s_0}$.

We also analyze the situations of $0^{-+}$ and $1^{++}$, and find that no matter what values of $M_B^2$ and $\sqrt{s_0}$ take, no optimal window for stable plateaus exist. That means the currents in Eqs.(\ref{J0-+}) and (\ref{J1++}) do not support the corresponding hexaquark molecular states.

\section{Decay analyses}\label{Decay}
In Refs. \cite{Qiao:2005av, Qiao:2007ce}, the heavy baryonium was introduced to explain the production and decays of $Y(4260)$, however, a decade has passed, the signal for heavy baryonium is still unclear. To finally ascertain these hidden charm and bottom baryonium, the straightforward procedure is to reconstruct them from its decay products, though the detailed characters of them still ask for more exploration. In Ref.\cite{Qiao:2005av} the hidden charm baryonium to $\pi^+\pi^-J/\psi$ decay was thought to be an easy-to-go process, where baryonium mass is below the threshold, while the $D\bar{D}$ exclusive process is almost impossible and the final states with strangeness should be suppressed. According this analysis, the heavy baryonia are above the $\Lambda_c\bar{\Lambda}_c$ threshold, and in this case the open-charm decay process of $\Lambda_c\bar{\Lambda}_c$ would be another potential choice. Analogously, the primary decay mode of b-sector heavy baryonium should be to $\Lambda_b\bar{\Lambda}_b$ process, as well the $\pi^+\pi^-\Upsilon$. The typical decay modes of the heavy baryonium for different quantum numbers are given in Table \ref{decay-mode}, and these processes are expected to be measurable in the running experiments like the BES III and LHCb.

\begin{table}[htbp]
\begin{center}\caption{Typical decay modes of the heavy baryomium for each quantum number, where $X$ denotes the baryonium respectively.}
\begin{tabular}{|c|c|c|}
\hline
\hline
$J^{PC}$       &  c-sector                                                              & b-sector   \\
\hline
$0^{++}$       & $X \to\Lambda_c\bar{\Lambda}_c$                   & $X \to\Lambda_b\bar{\Lambda}_b$ \\
\hline
$1^{--}$        & $X \to\Lambda_c\bar{\Lambda}_c$                    & $X \to\Lambda_b\bar{\Lambda}_b$ \\
                      & $X \to\pi^+\pi^-J/\psi$                                          & $X \to\pi^+\pi^-\Upsilon$ \\
\hline
\hline
\end{tabular}
\end{center}
\label{decay-mode}
\end{table}

\section{Conclusions}

In summary, we investigate in this work the hidden-bottom and -charm hexaquark structures in molecular configuration with quantum numbers $J^{PC }= 0^{++}$, $0^{-+}$, $1^{++}$ and $1^{--}$, in the framework of QCD sum rules. After constructing the appropriate interpreting currents, we perform the QCD sum rules analysis, where the vacuum condensates are considered up to dimension 12. Our results indicate that there exist two possible $\Lambda_b$-$\bar{\Lambda}_b$-like baryonium states with masses $11.84 \pm 0.22$ GeV and $11.72 \pm 0.26$ GeV for $0^{++}$  and $1^{--}$, respectively. The masses of their hidden-charm partners are found to be $5.19 \pm 0.24$ GeV and $4.78 \pm 0.23$ GeV, respectively. It is found the $0^{-+}$ and $1^{++}$ currents do not yield hadronic structures in any reasonable magnitudes of $M_B^2$ and $\sqrt{s_0}$, say no optimal window for stable plateaus exist. Results indicate that the $Y(4660)$ reported by Belle Collaboration \cite{Pakhlova:2008vn,Wang:2007ea} is close in magnitude to our calculation of $\Lambda_c$-$\bar{\Lambda}_c$ baryonium state. Moreover, the primary and potential decay modes of baryonium are analyzed, which might serve as a guide for experimental exploration.

It should be noted that the hidden-charm baryonium was once investigated by Chen, {\it et al.} \cite{Chen:2016ymy}. Analytically we can find an agreement with them, however, there exist two main differences between these two analyses. One is the threshold truncation of non-perturbative condensite, in our calculation the higher dimensional condensate $\langle \bar{q} q\rangle^2\langle G^2 \rangle$ is taken into account, whereas neglected in \cite{Chen:2016ymy}, which turns to be nonnegligible in $0^{-+}$ and $1^{++}$ states analyses. Another difference lies in the choice of continuum threshold $s_0$ and Borel paramter $M_B^2$. In the end the day, we find hidden charm baryonium states $0^{++}$ and $1^{--}$ might exist, however according to \cite{Chen:2016ymy} the existing ones with quantum numbers $0^{++}$ and $1^{++}$.

It it interesting to note that according to QCD sum rule analysis the $1^{--}$ hidden charm baryonium state lies above the open charm threshold, whereas it might be below the threshold in heavy quark chiral theory, which deserves further investigations. Moreover, we  notice that our result about the $1^{--}$ $\Lambda_c-\bar{\Lambda}_c$ baryonium state is in consistent with the large-$N_c$ QCD result of Ref.\cite{Liu:2008}.

\vspace{.7cm} {\bf Acknowledgments} \vspace{.0cm}

This work was supported in part by the Ministry of Science and Technology of the Peoples’ Republic of China(2015CB856703); by the National Natural Science Foundation of China(NSFC) under the Grants 11975236, 11635009, 11375200 and 11605039; by the Natural Science Foundation of Hebei Province with contract No. A2017205124, and by the Science Foundation of Hebei Normal University under Contract No. L2016B08.


\begin{widetext}
\appendix

\section{The spectral densities of $0^{++}$ $\Lambda_Q$-$\bar{\Lambda}_Q$ hexaquark states}

The $0^{++}$ $\Lambda_c$-$\bar{\Lambda}_c$ state spectral densities on the OPE side:
\begin{eqnarray}
\rho^{pert}(s)&=&\frac{1}{\pi^{10}}\int_{\alpha_{min}}^{\alpha_{max}} d \alpha \int_{\beta_{min}}^{1-\alpha} d \beta \bigg{\{}\frac{{\cal F}_{\alpha \beta}^{7}(-1+\alpha+\beta)^{4}}{3\times5\times7\times2^{19} \alpha^{6} \beta^{6}}+\frac{{\cal F}_{\alpha \beta}^{6} m_Q^{2}(1-\alpha-\beta)^{5}}{3\times5^2\times2^{19}  \alpha^{6} \beta^{6}}\bigg{\}}\; ,\\
\rho^{\langle G^2 \rangle}(s)&=&\frac{\langle G^2 \rangle}{\pi^{10}}\int_{\alpha_{min}}^{\alpha_{max}} d \alpha \int_{\beta_{min}}^{1-\alpha} d \beta \bigg{\{}\frac{{\cal F}_{\alpha \beta}^{4} (-1+\alpha+\beta)^{2} \left(3{\cal F}_{\alpha \beta}-5 m_Q^{2}(-1+\alpha+\beta)\right)}{3\times5\times2^{19} \alpha^{4} \beta^{4}}\nonumber\\
&+&\frac{{\cal F}_{\alpha \beta}^{3}m_Q^{2}(-1+\alpha+\beta)^{4}}{5\times3^2\times2^{21} \alpha^{6} \beta^{6}}\left(-4 m_Q^{2}\left(\alpha^{4}+\alpha^{3}(-1+\beta)+\alpha \beta^{3}+(-1+\beta) \beta^{3}\right)\right. \nonumber\\
&+&{\cal F}_{\alpha \beta}\left(2 \alpha^{3}-3 \alpha^{2}(-1+\beta)-3 \alpha \beta^{2}+\beta^{2}(3+2 \beta)) \right)\bigg{\}}\; ,\\
\rho^{\langle \bar{q} q\rangle^2}(s)&=&\frac{\langle \bar{q} q\rangle^2}{3\times2^{10}\pi^6}\int_{\alpha_{min}}^{\alpha_{max}} d \alpha \int_{\beta_{min}}^{1-\alpha} d \beta \bigg{\{}-\frac{{\cal F}_{\alpha \beta}^{4}(1-\alpha-\beta)}{\alpha^{3} \beta^{3}}-\frac{2{\cal F}_{\alpha \beta}^{3} m_Q^{2}(1-\alpha-\beta)^{2}}{\alpha^{3} \beta^{3}}\bigg{\}}\; ,\\
\rho^{\langle G^3 \rangle}(s) &=&\frac{\langle G^3 \rangle}{5\times3^2\times2^{22}\pi^{10}}\int_{\alpha_{min}}^{\alpha_{max}} d \alpha \int_{\beta_{min}}^{1-\alpha} d \beta {\cal F}_{\alpha \beta}^2  (\alpha +\beta -1)^4  \nonumber\\
&\times& \frac{5 {\cal F}_{\alpha \beta}^2+8 {\cal F}_{\alpha \beta} m_Q^2 (2 \alpha -3 \beta +3)-24 \alpha  m_Q^4 (\alpha +\beta -1)}{ \alpha ^3 \beta ^6}\; ,\\
\rho^{\langle \bar{q} q \rangle \langle \bar{q} G q \rangle}(s)&=&\frac{\langle \bar{q} q \rangle \langle \bar{q} G q \rangle}{3\times2^9\pi^6}\int_{\alpha_{min}}^{\alpha_{max}} d \alpha \int_{\beta_{min}}^{1-\alpha} d \beta -\frac{{\cal F}_{\alpha \beta}^{3}-3 m_Q^{2}{\cal F}_{\alpha \beta}^2(-1+\alpha+\beta)}{ \alpha^{2} \beta^{2}}\; ,\\
\rho^{\langle \bar{q} G q \rangle^2}&=&\frac{\langle \bar{q} G q \rangle^2}{2^{11}\pi^6}\int_{\alpha_{min}}^{\alpha_{max}} d \alpha \bigg{\{} -\frac{{\cal H}_\alpha^{2} }{2 (1-\alpha) \alpha}-\int_{\beta_{min}}^{1-\alpha} d \beta \frac{{\cal F}_{\alpha \beta} m_Q^{2} }{ \alpha \beta}\bigg{\}} \; ,\\
\rho^{\langle \bar{q} q\rangle^2\langle G^2 \rangle}&=&\frac{\langle \bar{q} q\rangle^2\langle G^2 \rangle}{3\times2^{11}\pi^6}\int_{\alpha_{min}}^{\alpha_{max}} d \alpha \bigg{\{}-\frac{{\cal H}_\alpha^{2}}{2 (1-\alpha) \alpha}-\int_{\beta_{min}}^{1-\alpha} d \beta\bigg[ \frac{{\cal F}_{\alpha \beta} m_Q^{2}}{ \alpha \beta}\nonumber\\
&-&\frac{1}{3 \alpha^{3} \beta^{3}}(-1+\alpha+\beta) \big[{\cal F}_{\alpha \beta} m_Q^{2}\left(\alpha^{3}+3 \alpha^{2}(-1+\beta)+3 \alpha \beta^{2}+(-3+\beta) \beta^{2}\right) \nonumber\\
&+&m_Q^{4}\left(\alpha^{4}+\alpha^{3}(-1+\beta)+\alpha \beta^{3}+(-1+\beta) \beta^{3}\right)\big]\bigg]\bigg{\}}\; ,\\
\rho^{\langle \bar{q} q\rangle^4}&=&\int_{\alpha_{min}}^{\alpha_{max}} d \alpha \frac{{\cal H}_\alpha}{48 \pi^{2}}\langle \bar{q} q\rangle^4\;,
\end{eqnarray}
where $M_B$ is the Borel parameter introduced by the Borel
transformation, $Q = c$ or $b$. Here, we also have the following definitions:
\begin{eqnarray}
{\cal F}_{\alpha \beta} &=& (\alpha + \beta) m_Q^2 - \alpha \beta s \; , {\cal H}_\alpha  = m_Q^2 - \alpha (1 - \alpha) s \; , \\
\alpha_{min} &=& \left(1 - \sqrt{1 - 4 m_Q^2/s} \right) / 2, \; , \alpha_{max} = \left(1 + \sqrt{1 - 4 m_Q^2 / s} \right) / 2  \; , \\
\beta_{min} &=& \alpha m_Q^2 /(s \alpha - m_Q^2).
\end{eqnarray}

\section{The spectral densities of $0^{-+}$ $\Lambda_Q$-$\bar{\Lambda}_Q$ hexaquark states}

The $0^{-+}$ $\Lambda_c$-$\bar{\Lambda}_c$ state spectral densities on the OPE side:
\begin{eqnarray}
\rho^{pert}(s)&=&\frac{1}{\pi^{10}}\int_{\alpha_{min}}^{\alpha_{max}} d \alpha \int_{\beta_{min}}^{1-\alpha} d \beta \bigg{\{}\frac{{\cal F}_{\alpha \beta}^{7}(-1+\alpha+\beta)^{4}}{3\times5\times7\times2^{19} \alpha^{6} \beta^{6}}+\frac{{\cal F}_{\alpha \beta}^{6} m_Q^{2}(\alpha+\beta-1)^{5}}{3\times5^2\times2^{19}  \alpha^{6} \beta^{6}} \bigg{\}}\; ,\\
\rho^{\langle G^2 \rangle}(s)&=&\frac{\langle G^2 \rangle}{\pi^{10}}\int_{\alpha_{min}}^{\alpha_{max}} d \alpha \int_{\beta_{min}}^{1-\alpha} d \beta \bigg{\{}\frac{{\cal F}_{\alpha \beta}^{4} (-1+\alpha+\beta)^{2} \left(-3{\cal F}_{\alpha \beta}-5 m_Q^{2}(-1+\alpha+\beta)\right)}{3\times5\times2^{19} \alpha^{4} \beta^{4}}\nonumber\\
&-&\frac{{\cal F}_{\alpha \beta}^{3}m_Q^{2}(-1+\alpha+\beta)^{4}}{5\times3^2\times2^{21} \alpha^{6} \beta^{6}}\left(4 m_Q^{2}\left(\alpha^{4}+\alpha^{3}(-1+\beta)+\alpha \beta^{3}+(-1+\beta) \beta^{3}\right)\right. \nonumber\\
&+&{\cal F}_{\alpha \beta}\left(2 \alpha^{3}-3 \alpha^{2}(-1+\beta)-3 \alpha \beta^{2}+\beta^{2}(3+2 \beta)) \right)\bigg{\}}\; ,\\
\rho^{\langle \bar{q} q\rangle^2}(s)&=&\frac{\langle \bar{q} q\rangle^2}{3\times2^{10}\pi^6}\int_{\alpha_{min}}^{\alpha_{max}} d \alpha \int_{\beta_{min}}^{1-\alpha} d \beta \bigg{\{}\frac{{\cal F}_{\alpha \beta}^{4}(1-\alpha-\beta)}{\alpha^{3} \beta^{3}}-\frac{2{\cal F}_{\alpha \beta}^{3} m_Q^{2}(1-\alpha-\beta)^{2}}{\alpha^{3} \beta^{3}}\bigg{\}}\; ,\\
\rho^{\langle G^3 \rangle}(s) &=&\frac{\langle G^3 \rangle}{5\times3^2\times2^{22}\pi^{10}}\int_{\alpha_{min}}^{\alpha_{max}} d \alpha \int_{\beta_{min}}^{1-\alpha} d \beta {\cal F}_{\alpha \beta}^2  (\alpha +\beta -1)^4  \nonumber\\
&\times& \frac{-5 {\cal F}_{\alpha \beta}^2-8 {\cal F}_{\alpha \beta} m_Q^2 (8 \alpha +3 \beta -3)-24 \alpha  m_Q^4 (\alpha +\beta -1)}{ \alpha ^3 \beta ^6}\; ,\\
\rho^{\langle \bar{q} q \rangle \langle \bar{q} G q \rangle}(s)&=&\frac{\langle \bar{q} q \rangle \langle \bar{q} G q \rangle}{3\times2^9\pi^6}\int_{\alpha_{min}}^{\alpha_{max}} d \alpha \int_{\beta_{min}}^{1-\alpha} d \beta \frac{{\cal F}_{\alpha \beta}^{3}+3 m_Q^{2}{\cal F}_{\alpha \beta}^2(-1+\alpha+\beta)}{ \alpha^{2} \beta^{2}}\; ,\\
\rho^{\langle \bar{q} G q \rangle^2}&=&\frac{\langle \bar{q} G q \rangle^2}{2^{11}\pi^6}\int_{\alpha_{min}}^{\alpha_{max}} d \alpha \bigg{\{} \frac{{\cal H}_\alpha^{2} }{2 (1-\alpha) \alpha}-\int_{\beta_{min}}^{1-\alpha} d \beta \frac{{\cal F}_{\alpha \beta} m_Q^{2} }{ \alpha \beta}\bigg{\}} \; ,\\
\rho^{\langle \bar{q} q\rangle^2\langle G^2 \rangle}&=&\frac{\langle \bar{q} q\rangle^2\langle G^2 \rangle}{3\times2^{11}\pi^6}\int_{\alpha_{min}}^{\alpha_{max}} d \alpha \bigg{\{}\frac{{\cal H}_\alpha^{2}}{2 (1-\alpha) \alpha}-\int_{\beta_{min}}^{1-\alpha} d \beta\bigg[ \frac{{\cal F}_{\alpha \beta} m_Q^{2}}{ \alpha \beta}\nonumber\\
&+&\frac{1}{3 \alpha^{3} \beta^{3}}(-1+\alpha+\beta) \big[{\cal F}_{\alpha \beta} m_Q^{2}\left(\alpha^{3}+3 \alpha^{2}(-1+\beta)+3 \alpha \beta^{2}+(-3+\beta) \beta^{2}\right) \nonumber\\
&+&m_Q^{4}\left(\alpha^{4}+\alpha^{3}(-1+\beta)+\alpha \beta^{3}+(-1+\beta) \beta^{3}\right)\big]\bigg]\bigg{\}}\; ,\\
\rho^{\langle \bar{q} q\rangle^4}&=&\int_{\alpha_{min}}^{\alpha_{max}} d \alpha \frac{2m_Q^2-3{\cal H}_\alpha}{144 \pi^{2}}\langle \bar{q} q\rangle^4\;.
\end{eqnarray}

\section{The spectral densities of $1^{--}$ $\Lambda_Q$-$\bar{\Lambda}_Q$ hexaquark states}

The $1^{--}$ $\Lambda_c$-$\bar{\Lambda}_c$ state spectral densities on the OPE side:

\begin{eqnarray}
\rho^{pert} (s) &=& \frac{1}{\pi^{10}} \int^{\alpha_{max}}_{\alpha_{min}} d \alpha \int^{1 - \alpha}_{\beta_{min}} d \beta  \frac{7{\cal F}^6_{\alpha \beta} (\alpha + \beta - 1)^5 m_Q^2+ {\cal F}_{\alpha \beta}^7  (\alpha + \beta - 1)^4 (\alpha + \beta + 4 ) }{3\times 7 \times 5^2 \times 2^{19} \alpha^6 \beta^6}\;,\\
\rho^{\langle G^2 \rangle}(s) &=& \frac{\langle g_s^2 G^2\rangle}{\pi^{10}} \int^{\alpha_{max}}_{\alpha_{min}} d \alpha \int^{1 - \alpha}_{\beta_{min}} d \beta \bigg{\{} \frac{{\cal F}_{\alpha \beta}^5 (\alpha + \beta -1)^2 (\alpha + \beta +2)}{3\times5\times2^{19}\alpha^4\beta^4}+\frac{m_Q^2 {\cal F}_{\alpha \beta}^4 (\alpha + \beta -1)^3}{3\times2^{19}\alpha^4\beta^4}\nonumber\\
&+&\frac{{\cal F}_{\alpha \beta}^3 m_Q^2 (\alpha + \beta -1)^4 }{3^2\times5\times2^{21}\alpha^6\beta^6} \bigg [ 4 m_Q^2 \bigg(\alpha^4 +\alpha^3(\beta-1) +\alpha\beta^3+\beta^3(\beta-1)\bigg)\nonumber\\
&+&{\cal F}_{\alpha \beta}\bigg( \alpha^4 +3 \alpha^2 (\beta-1) +\alpha \beta^2 (\beta+3) +\alpha^3(\beta+7) +\beta^2(\beta^2+7\beta-3) \bigg) \bigg] \bigg{\}}\;,\\
\rho^{\langle \bar{q} q \rangle^2}(s)&=& \frac{\langle \bar{q} q \rangle^2}{3\times2^{11}\pi^6} \int^{\alpha_{max}}_{\alpha_{min}} d \alpha \int^{1 - \alpha}_{\beta_{min}} d \beta \frac{4 m_Q^2 {\cal F}_{\alpha \beta}^3 (\alpha+\beta-1)^2+{\cal F}_{\alpha \beta}^4 [(\alpha+\beta)^2 - 1]}{\alpha^3\beta^3}\;,\\
\rho^{\langle G^3 \rangle}(s) &=&\frac{\langle G^3 \rangle}{5\times3^2\times2^{22}\pi^{10}}\int_{\alpha_{min}}^{\alpha_{max}} d \alpha \int_{\beta_{min}}^{1-\alpha} d \beta {\cal F}_{\alpha \beta}^2  (\alpha +\beta -1)^4  \nonumber\\
&\times& \frac{(4+\alpha+\beta) {\cal F}_{\alpha \beta}^2+8 {\cal F}_{\alpha \beta} m_Q^2 ( \alpha^2 +3 (\beta-1) +\alpha(7+\beta))+24 \alpha  m_Q^4 (\alpha +\beta -1)}{ \alpha ^3 \beta ^6}\; ,\\
\rho^{\langle \bar{q} q \rangle \langle \bar{q} G q \rangle}(s)&=&\frac{\langle \bar{q} q \rangle \langle \bar{q} G q \rangle}{3\times2^9\pi^6}\int_{\alpha_{min}}^{\alpha_{max}} d \alpha \int_{\beta_{min}}^{1-\alpha} d \beta -\frac{{\cal F}_{\alpha \beta}^{3}(\alpha+\beta)+3 m_Q^{2}{\cal F}_{\alpha \beta}^2(-1+\alpha+\beta)}{ \alpha^{2} \beta^{2}}\; ,\\
\rho^{\langle \bar{q} G q \rangle^2}&=&\frac{\langle \bar{q} G q \rangle^2}{2^{11}\pi^6}\int_{\alpha_{min}}^{\alpha_{max}} d \alpha \bigg{\{} -\frac{{\cal H}_\alpha^{2} }{2 (1-\alpha) \alpha}+\int_{\beta_{min}}^{1-\alpha} d \beta \bigg[\frac{{\cal F}_{\alpha \beta} m_Q^{2} }{ \alpha \beta}+\frac{{\cal F}_{\alpha \beta}^2 }{ 2\alpha \beta}\bigg]\bigg{\}} \; ,\\
\rho^{\langle \bar{q} q\rangle^2\langle G^2 \rangle}&=&\frac{\langle \bar{q} q\rangle^2\langle G^2 \rangle}{3\times2^{11}\pi^6}\int_{\alpha_{min}}^{\alpha_{max}} d \alpha \bigg{\{}-\frac{{\cal H}_\alpha^{2}}{2 (1-\alpha) \alpha}+\int_{\beta_{min}}^{1-\alpha} d \beta\bigg[ \frac{{\cal F}_{\alpha \beta} m_Q^{2}}{ \alpha \beta}+ \frac{{\cal F}_{\alpha \beta}^2}{2 \alpha \beta}\nonumber\\
&-&\frac{1}{3 \alpha^{3} \beta^{3}}(-1+\alpha+\beta) \big[m_Q^{4}\left(\alpha^{4}+\alpha^{3}(-1+\beta)+\alpha \beta^{3}+(-1+\beta) \beta^{3}\right)+{\cal F}_{\alpha \beta} m_Q^{2}\nonumber\\
&\times&\left(\alpha^{4}+3 \alpha^{2}(-1+\beta)+(3+\beta) \alpha \beta^{2}+\alpha ^3 (\beta +4)+\beta ^2 (\beta ^2+4 \beta -3 ) \right)\big]\bigg]\bigg{\}}\; ,\\
\rho^{\langle \bar{q} q\rangle^4}&=&\int_{\alpha_{min}}^{\alpha_{max}} d \alpha \frac{{\cal H}_\alpha-m_Q^2}{72 \pi^{2}}\langle \bar{q} q\rangle^4\;.
\end{eqnarray}

\section{The spectral densities of $1^{++}$ $\Lambda_Q$-$\bar{\Lambda}_Q$ hexaquark states}

The $1^{++}$ $\Lambda_c$-$\bar{\Lambda}_c$ state spectral densities on the OPE side:

\begin{eqnarray}
\rho^{pert} (s) &=& \frac{1}{\pi^{10}} \int^{\alpha_{max}}_{\alpha_{min}} d \alpha \int^{1 - \alpha}_{\beta_{min}} d \beta  \frac{{\cal F}_{\alpha \beta}^7  (\alpha + \beta - 1)^4 (\alpha + \beta + 4 )-{\cal F}^6_{\alpha \beta} (\alpha + \beta - 1)^5 m_Q^2 }{3 \times 7\times 5^2 \times 2^{19} \alpha^6 \beta^6}\;,\\
\rho^{\langle G^2 \rangle}(s) &=& \frac{\langle g_s^2 G^2\rangle}{\pi^{10}} \int^{\alpha_{max}}_{\alpha_{min}} d \alpha \int^{1 - \alpha}_{\beta_{min}} d \beta \bigg{\{} \frac{{\cal F}_{\alpha \beta}^5 (\alpha + \beta -1)^2 (\alpha + \beta +2)}{3\times5\times2^{19}\alpha^4\beta^4}-\frac{m_Q^2 {\cal F}_{\alpha \beta}^4 (\alpha + \beta -1)^3}{3\times2^{19}\alpha^4\beta^4}\nonumber\\
&+&\frac{{\cal F}_{\alpha \beta}^3 m_Q^2 (\alpha + \beta -1)^4 }{3^2\times5\times2^{21}\alpha^6\beta^6} \bigg [ -4 m_Q^2 \bigg(\alpha^4 +\alpha^3(\beta-1) +\alpha\beta^3+\beta^3(\beta-1)\bigg)\nonumber\\
&+&{\cal F}_{\alpha \beta}\bigg( \alpha^4 +3 \alpha^2 (\beta-1) +\alpha \beta^2 (\beta-3) +\alpha^3(\beta+1) +\beta^2(\beta^2+\beta+3) \bigg) \bigg] \bigg{\}}\;,\\
\rho^{\langle \bar{q} q \rangle^2}(s)&=& \frac{\langle \bar{q} q \rangle^2}{3\times2^{11}\pi^6} \int^{\alpha_{max}}_{\alpha_{min}} d \alpha \int^{1 - \alpha}_{\beta_{min}} d \beta \frac{-4 m_Q^2 {\cal F}_{\alpha \beta}^3 (\alpha+\beta-1)^2+{\cal F}_{\alpha \beta}^4 [(\alpha+\beta)^2 - 1]}{\alpha^3\beta^3}\;,\\
\rho^{\langle G^3 \rangle}(s) &=&\frac{\langle G^3 \rangle}{5\times3^2\times2^{22}\pi^{10}}\int_{\alpha_{min}}^{\alpha_{max}} d \alpha \int_{\beta_{min}}^{1-\alpha} d \beta {\cal F}_{\alpha \beta}^2  (\alpha +\beta -1)^4  \nonumber\\
&\times& \frac{(4+\alpha+\beta) {\cal F}_{\alpha \beta}^2+8 {\cal F}_{\alpha \beta} m_Q^2 ( \alpha^2 +3 (1-\beta) +\alpha(1+\beta))-24 \alpha  m_Q^4 (\alpha +\beta -1)}{ \alpha ^3 \beta ^6}\; ,\\
\rho^{\langle \bar{q} q \rangle \langle \bar{q} G q \rangle}(s)&=&\frac{\langle \bar{q} q \rangle \langle \bar{q} G q \rangle}{3\times2^9\pi^6}\int_{\alpha_{min}}^{\alpha_{max}} d \alpha \int_{\beta_{min}}^{1-\alpha} d \beta -\frac{{\cal F}_{\alpha \beta}^{3}(\alpha+\beta)-3 m_Q^{2}{\cal F}_{\alpha \beta}^2(-1+\alpha+\beta)}{ \alpha^{2} \beta^{2}}\; ,\\
\rho^{\langle \bar{q} G q \rangle^2}&=&\frac{\langle \bar{q} G q \rangle^2}{2^{11}\pi^6}\int_{\alpha_{min}}^{\alpha_{max}} d \alpha \bigg{\{} -\frac{{\cal H}_\alpha^{2} }{2 (1-\alpha) \alpha}+\int_{\beta_{min}}^{1-\alpha} d \beta \bigg[-\frac{{\cal F}_{\alpha \beta} m_Q^{2} }{ \alpha \beta}+\frac{{\cal F}_{\alpha \beta}^2 }{ 2\alpha \beta}\bigg]\bigg{\}} \; ,\\
\rho^{\langle \bar{q} q\rangle^2\langle G^2 \rangle}&=&\frac{\langle \bar{q} q\rangle^2\langle G^2 \rangle}{3\times2^{11}\pi^6}\int_{\alpha_{min}}^{\alpha_{max}} d \alpha \bigg{\{}-\frac{{\cal H}_\alpha^{2}}{2 (1-\alpha) \alpha}+\int_{\beta_{min}}^{1-\alpha} d \beta\bigg[-\frac{{\cal F}_{\alpha \beta} m_Q^{2}}{ \alpha \beta}+ \frac{{\cal F}_{\alpha \beta}^2}{2 \alpha \beta}\nonumber\\
&-&\frac{1}{3 \alpha^{3} \beta^{3}}(-1+\alpha+\beta) \big[m_Q^{4}\left(\alpha^{4}+\alpha^{3}(-1+\beta)+\alpha \beta^{3}+(-1+\beta) \beta^{3}\right)+{\cal F}_{\alpha \beta} m_Q^{2}\nonumber\\
&\times&\left(\alpha^{4}+3 \alpha^{2}(-2+\beta)+(-3+\beta) \alpha \beta^{2}+\alpha ^3 (\beta -2)+\beta ^2 (\beta ^2-2 \beta -3 ) \right)\big]\bigg]\bigg{\}}\; ,\\
\rho^{\langle \bar{q} q\rangle^4}&=&\int_{\alpha_{min}}^{\alpha_{max}} d \alpha \frac{{\cal H}_\alpha}{72 \pi^{2}}\langle \bar{q} q\rangle^4\;.
\end{eqnarray}

\end{widetext}
\end{document}